\documentclass[prl, superscriptaddress, twocolumn, pdflatex]{revtex4}
\usepackage{amsmath, amsfonts, amssymb,graphicx, dcolumn}

\begin{document}
\title{Conditional preparation of states containing a definite number of photons}
\author{Malcolm N. O'Sullivan}
\email{osulliva@optics.rochester.edu}
\affiliation{The Institute
of Optics, University of Rochester, Rochester, NY 14627, USA}
\author{Kam Wai Clifford Chan}
\affiliation{The Institute of Optics, University of Rochester,
Rochester, NY 14627, USA}
\author{Vasudevan Lakshminarayanan}
\affiliation{School of Optometry, Department of Electrical and Computer Engineering, and Department of Physics and Astronomy, University of Waterloo,
Waterloo, ON, N2L 3G1, Canada}
\author{Robert W. Boyd}
\affiliation{The Institute of Optics, University of Rochester,
Rochester, NY 14627, USA}
\date{\today}

\begin{abstract}
A technique for conditionally creating single- or multimode
photon-number states is analyzed using Bayesian theory. We consider
the heralded $N$-photon states created from the photons produced by
an unseeded optical parametric amplifier when the heralding detector
is the time-multiplexed photon-number-resolving detector recently
demonstrated by Fitch, {\it et al}.~[Phys. Rev. A {\bf 68}, 043814
(2003).] and simultaneously by Achilles, {\it et al}.~[Opt. Lett.
{\bf 28}, 2387 (2003).]. We find that even with significant loss in
the heralding detector, fields with sub-Poissonian photon-number
distributions can be created.  We also show that heralded multimode
fields created using this technique are more robust against detector
loss than are single-mode fields.
\end{abstract}

\maketitle

A Fock state of the electromagnetic field contains a definite number of photons and hence displays no intensity fluctuations. In contrast, the distribution of the intensities of a classical field of light has a variance at least as large as its mean~\cite{MandelWolfPoisson}.  Thus, Fock states and other states with sub-Poissonian photon-number statistics are of interest not only because of their fundamentally nonclassical nature but also because of their potential to make radiometric measurements with greater accuracy than possible using classical states of light. Research in Fock state generation is also motivated by its applications in quantum information science~\cite{DrummondRMP94} including the areas of quantum cryptography~\cite{GisinRMP02} and quantum computing~\cite{KnillNature01}.

Unfortunately, it can be experimentally challenging to produce light fields that contain definite numbers of photons. Several theoretical proposals aimed at creating Fock states ``on demand'' have been offered~\cite{BrownPRA03, GeremiaPRL06, KurizkiPRA96}, but they involve trapped atoms in high-Q cavities and to the best of our knowledge only single-photon states have been created in the laboratory~\cite{KimbleSci04}.  Several stochastic sources of one- and two-photon states have been experimentally demonstrated using various processes including atomic and molecular fluorescence~\cite{DarquieSci05, LukishovaQE03}, Coulomb blockade for electrons~\cite{KimNat99} and action in a micromaser~\cite{VarcoeNat00}. In addition, a particularly simple method of generating single-photon states using the photon pairs created in the process of spontaneous parametric down-conversion was first demonstrated by Hong and Mandel~\cite{HongPRL85}.  In their experiment, they used the single-photon detection of one photon of the pair to ``herald'' the presence of the other photon of the pair.  This heralding scheme is generalizable to create multiphoton states of light. In an unseeded optical parametric amplifier (OPA), pump photons interact nonlinearly with a noncentrosymmetric crystal to produce multiple pairs of less energetic photons termed the signal and idler through the process of parametric down-conversion (PDC). In general, the number of photons found in the signal and idler fields is random and known to obey thermal statistics, which can be characterized by the gain $g$ of the OPA. But since the emission of every signal photon requires the simultaneous emission of an idler photon, if one knows the number of photons in the idler field then the number in the signal field is determined with complete certainty. Thus, in principle, by allowing transmission of the signal field only when the idler field is found to contain $n$ photons, the signal field will be guaranteed to also contain exactly $n$ photons. This idea was considered by Holmes {\it et al.}~\cite{HolmesPRA89} for ideal photon-number-resolving detection.

However, the photon-number-resolving detection needed to create heralded multiphoton states is not trivial to implement and is an area of active research.  Photon-number-resolving detection has been achieved using superconducting transition-edge sensors~\cite{MillerAPL03}, visible light counters~\cite{WaksQE03}, charge integration photon detectors~\cite{FujiwaraAPL05}, and by time-multiplexed detection involving only commercially available APDs~\cite{FitchPRA03, AchillesOL03}. In the present work, we show that, in particular, the photon-number resolution achieved using time-multiplexed detectors (TMDs), as first conceived by Fitch {\it et al.}~\cite{FitchPRA03} and independently by Achilles {\it et al.}~\cite{AchillesOL03}, is well-suited for creating heralded fields with near-definite numbers of photons even in the presence of significant detector loss.  This technique has been demonstrated experimentally for two-photon states~\cite{AchillesPRL06} but has not yet been analyzed in a comprehensive fashion for states containing greater numbers of photons.

\begin{figure}
    \includegraphics[scale=0.70]{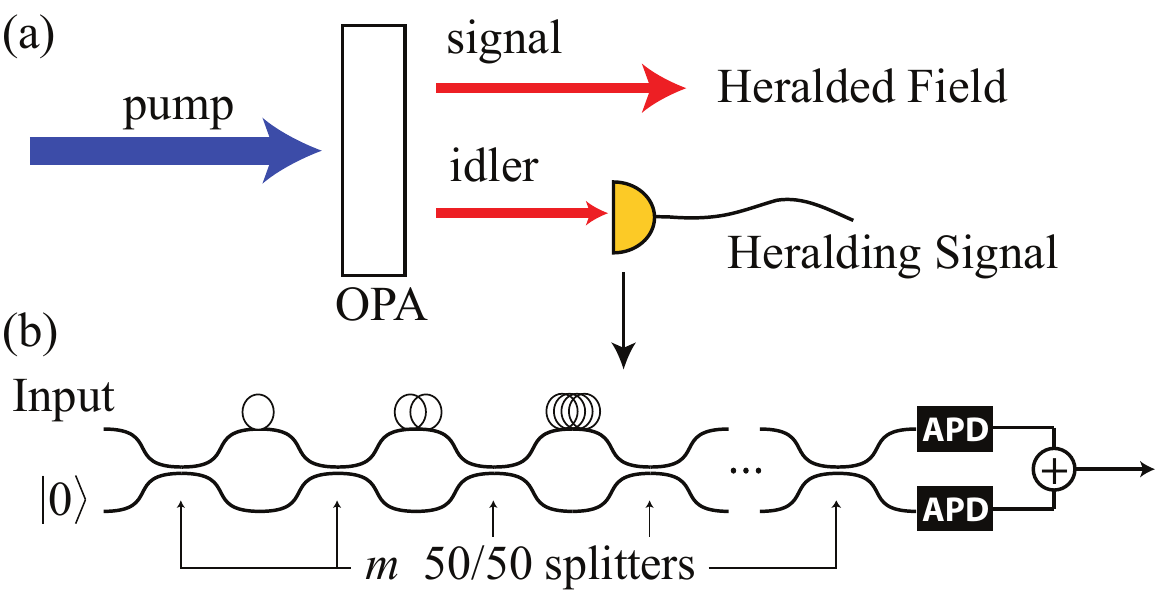}
    \caption{\label{fig-setup} (a) Scheme to produce heralded fields
    containing a definite number of photons.  An unseeded optical
    parametric amplifier (OPA) produces photon pairs through the
    process of parametric down-conversion (PDC). When a time-multiplexed photon-number-resolving
    detector (shown in (b)) detects $n_i$ photons in the idler field,
    it heralds the presence of $n_s$ photons in the signal field with
    probability $P_{\rm sig}(n_s|n_i)$.}
\end{figure}

A schematic of the proposed heralding technique is shown in Fig.~\ref{fig-setup}.  As mentioned above, the source of photon pairs is the output of an unseeded OPA and the heralding detector is a TMD. The TMD provides photon-number resolution by first randomly splitting the incoming state into many pathways that are separated in time from one another, i.e.~they are separated into time bins.  The number of photons in the incoming field can be estimated to be equal to the total number of photons detected by two avalanche photodiodes (APDs). Any errors in this estimation will introduce uncertainty into the number of photons contained in the heralded state. Undercounting is the main source of errors in TMDs. The two primary mechanisms for undercounting errors are losses and inadequate splitting of the incoming beam.  Losses in the detection system stemming both from the beam splitters and from inefficiencies in detectors themselves can be quite significant and must be included in the analysis. Inadequate splitting occurs because of the finite number of times the incoming beam is split. Hence, a chance exists that two or more photons will occupy the same time bin resulting in an undercounting error at the APDs, which can perform only on/off detection.  More quantitatively, a system containing $m$ beam splitters will have $M\equiv2^m$ different time bins.  If the number of incoming photons $N$ is much less than $M$, the probability of undercounting due to imperfect splitting will be small. However, when $N$ is comparable to $M$, we must account for splitting errors.  Using a standard urn model~\cite{Johnson2005, FitchPRA03}, we can calculate the conditional probability of detecting $n$ photons given that the incoming field has $N$ photons in it.  This probability is given by the expression
\begin{equation}
    \label{eq-pngN}
    P_{\rm det}(n|N) = \binom{M}{n} \sum_{j=0}^{n} (-1)^j
    \binom{n}{j} \left[ (1-\eta) + \frac{\eta(n-j)}{M}\right]^N,
\end{equation}
where $\eta$ is the single-photon detection efficiency. It is not difficult to verify that in the limit $M\gg 1$ and $M\gg N$ detection errors due to imperfect splitting are negligible and that $P_{\rm det}(n|N) \rightarrow \binom{N}{n}(1-\eta)^{N-n}\eta^n$, which is the expected distribution for an ideal photon-number-resolving detector subject to a loss of $1-\eta$ per photon. We note that detector dark counts are a source of overcounting errors.  However, in the present analysis, we do not include the effect of detector dark counts, since often they can be rendered neglible through detector gating or by use of detectors with intrinsically low dark count rates (commercially available APD dark count rates can be as low as 5 dark counts per second).

\begin{figure}
    \includegraphics[scale=0.6]{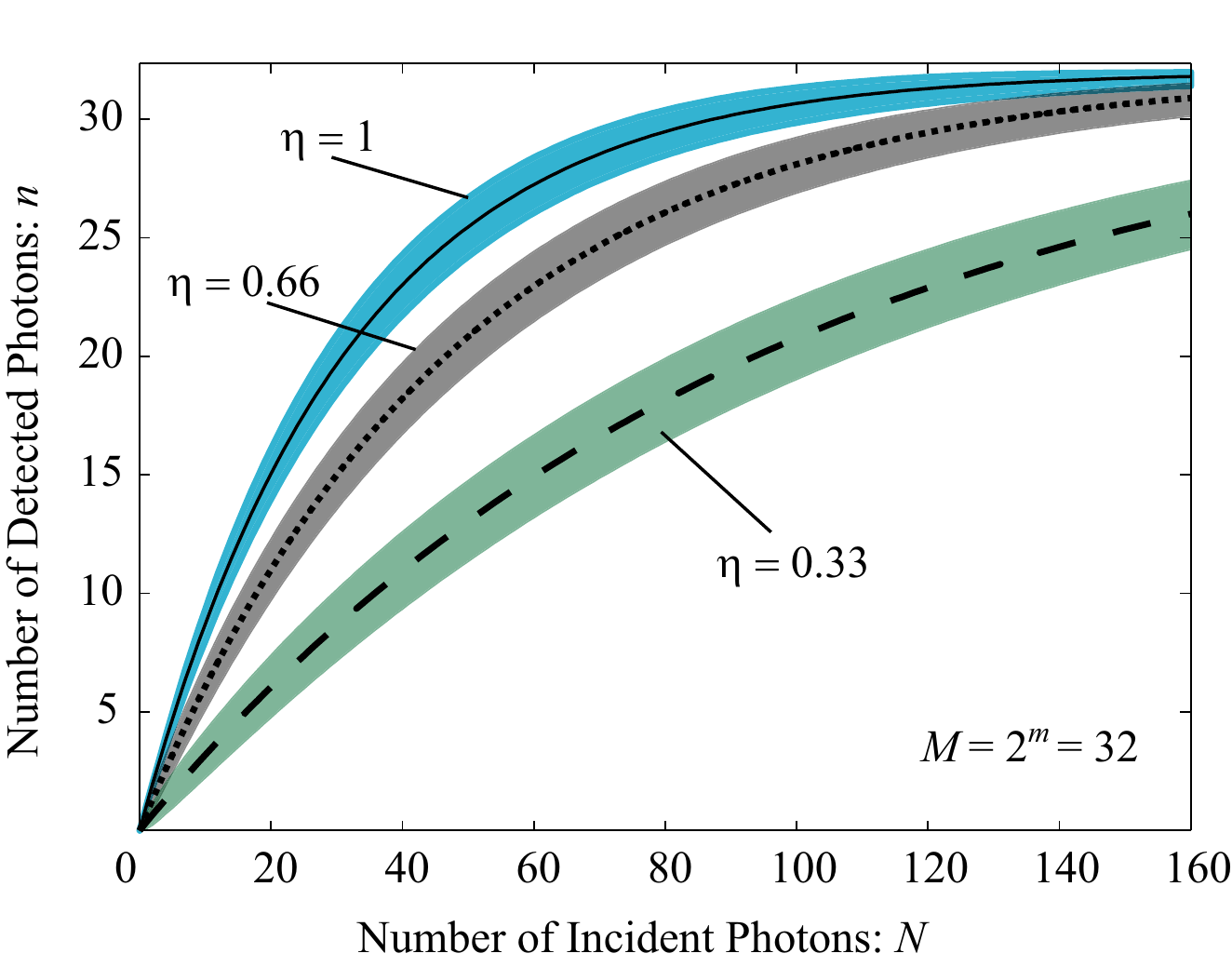}
    \caption{\label{fig-pnN} Regions where there is a significant probability of detecting $n$ photons conditioned on $N$ photons being incident on the TMD for three different single-photon detection efficiencies ($\eta =$ 1.0, 0.66, 0.33).  The mean number of photons detected given $N$ photons input into the detection system are plotted using the solid ($\eta = 1$), dotted ($\eta = 0.66$) and dashed ($\eta = 0.33$) lines.  The shaded regions are within one standard deviation of the means.}
\end{figure}

The response of the detection system is shown in Fig.~\ref{fig-pnN} for a system comprised of five beam splitters (M=32) for three different scenarios.  For no loss in the TMD ($\eta=1$) and few enough incident photons, the detector will accurately measure the number of photons in the field, i.e.~the initial slope of the mean of the probability distribution is linear with unity slope. Once the number of incident photons increases and becomes comparable to the maximum number of photons $M$ the system can count, undercounting will become very likely and the detection system will saturate.  As the loss in the system increases ($\eta$ decreases), two effects are seen: (1) the initial slope decreases and is given roughly by $\eta$ and (2) the error in the estimate increases as seen by the increase in the vertical spread of the distribution in the plot.

As we already mentioned, in a heralding system any error in the detection of the number of photons in the heralding field will directly lead to an uncertainty in the number of the photons in the heralded field.  If this uncertainty, as determined by the variance of the number photons estimated to be in the heralding field, becomes larger than the mean number of photons in the heralded field, the state can, in principle, be produced by classical means. However, since the source of the photon pairs is an OPA, we can use this {\it a priori} information about its photon emission statistics to improve our estimation through the use of Bayesian theory.

The probability that an OPA characterized by gain $g$ will emit $k$ pairs of photons is given by  $P_{\rm OPA}(k) = \left| \langle k,k|\psi\rangle\right|^2$, where $|k,k\rangle$ is the state containing $k$ photons in both the signal and idler fields and $|\psi\rangle$ is the state emitted by the OPA, which is given by $\exp(g\hat{a}^\dag_i\hat{a}^\dag_s-g\hat{a}_i\hat{a}_s)|0,0\rangle$ \cite{WallsMilburn95}. As a notational device, we will identify the signal field with the heralded field and the idler field with the heralding field.  It can be shown that the resulting photon-number distribution for either the signal or idler field obey Bose-Einstein statistics and is given by
\begin{equation}
    P_{\rm OPA}(k) = (1-\tanh^2\!g)\tanh^{2k}\!g.
\end{equation}
We can calculate the resulting mixed state of the signal field given that $n_i$ photons are detected in the idler field. The density matrix is given by
\begin{equation}
    \label{eq-rhosig}
    \hat{\rho}_{\rm sig}(n_{\rm i}) = \sum_{n_{\rm s}=n_{\rm i}}^\infty P_{\rm
    sig}(n_{\rm s}|n_{\rm i})\:|n_{\rm s}\rangle\langle n_{\rm s}|,
\end{equation}
where $|n_{\rm s}\rangle$ is a Fock state of the signal field with $n_{\rm s}$ photons in it. The function $P_{\rm sig}(n_{\rm s}|n_{\rm i})$ is the probability that the signal field has $n_{\rm s}$ photons in it given that $n_{\rm i}$ photons are detected in the idler field.  According to Bayes' theorem, it is given by the expression
\begin{equation}
    \label{eq-pNgn}
     P_{\rm sig}(n_{\rm s}|n_{\rm i}) = \frac{P_{\rm det}(n_{\rm i}|n_{\rm s})P_{\rm OPA}(n_{\rm s})}{\sum_{k=0}^\infty P_{\rm det}(n_{\rm i}|k)\,P_{\rm OPA}(k)}.
\end{equation}

We estimate the number of photons in the signal field using maximum-likelihood estimation.  The maximum-likelihood estimate of the number of photons in the signal field given $n_{\rm i}$ photons detected in the idler field is found using
\begin{equation}
    \label{eq-Nml}
    n_{\rm s}^{\rm ml}(n_{\rm i}) = \arg\max_{n_{\rm s}} P_{\rm sig}(n_{\rm s}|n_{\rm i}),
\end{equation}
where the $\arg\max$ function returns the value of $n_{\rm s} \in \{n_{\rm i},n_{\rm i}+1,\ldots,\infty\}$ that maximizes $P_{\rm sig}(n_{\rm s}|n_{\rm i})$.

\begin{figure}
    \includegraphics[scale=0.65]{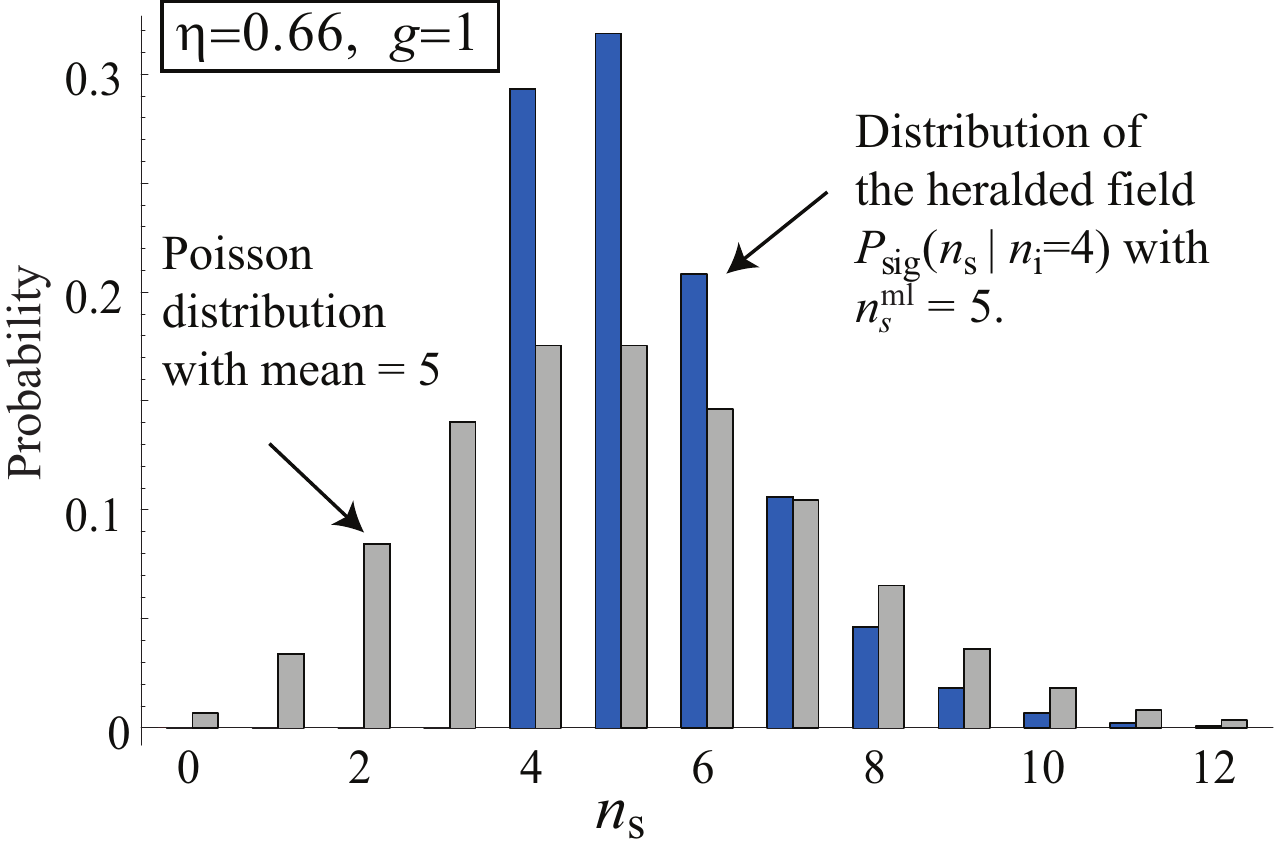}
    \caption{\label{fig-psig4} Comparison of the photon-number statistics for a typical heralded field with $n_{\rm s}^{\rm ml}=5$ (in blue) and a Poisson-distributed field with a mean photon number of 5 (in gray). The heralded field, for the case in which $\eta = 0.66$ and $g=1$, is conditioned on the detection of 4 photons in the idler field by the TMD. }
\end{figure}
\begin{figure}
    \includegraphics[scale=0.6]{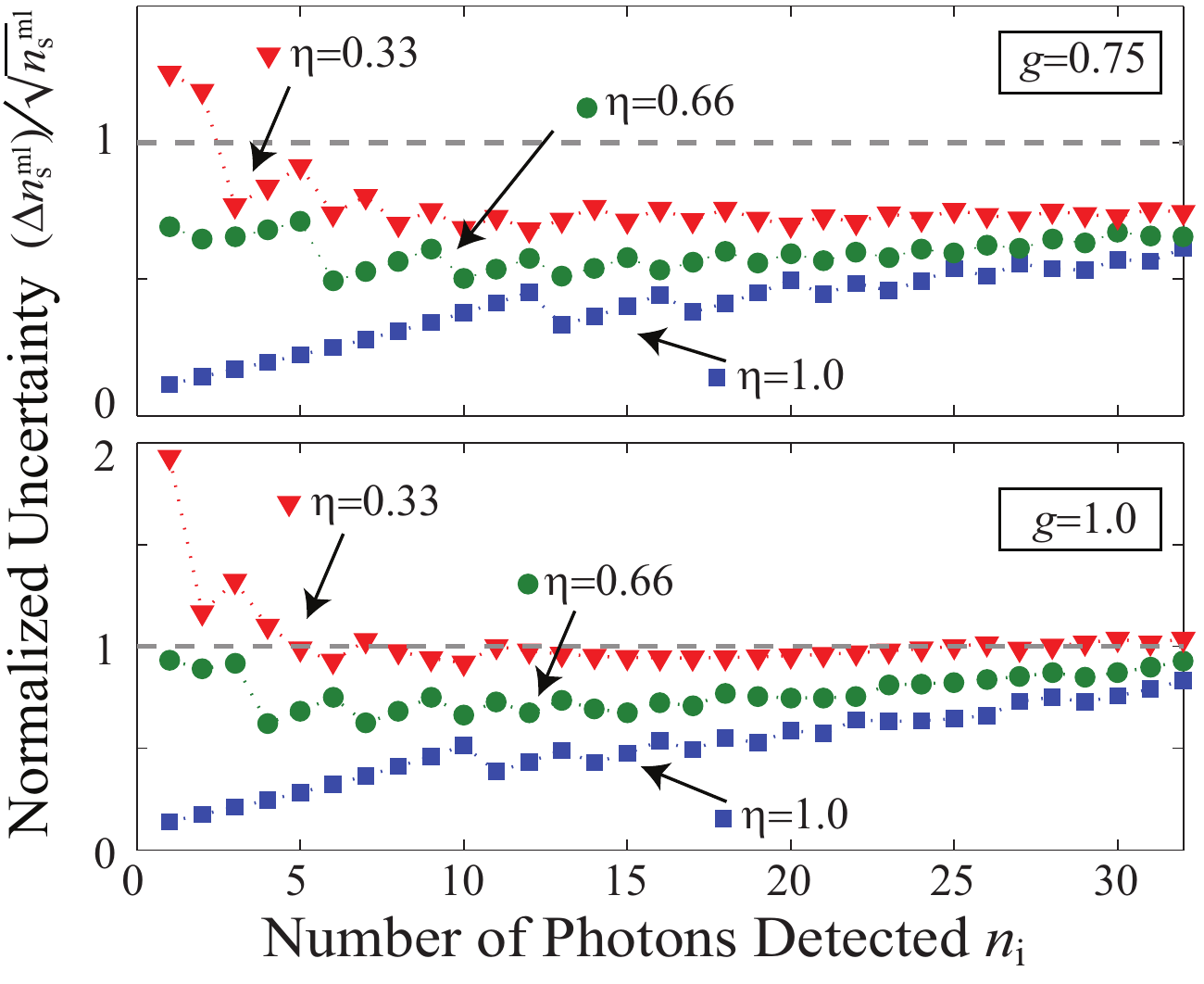}
    \caption{\label{fig-relerr} Uncertainty in the number of photons in a heralded field when $n_{\rm i}$ photons are detected in the heralding field normalized to the uncertainty in a comparable Poisson-distributed field with mean number of photons $n_{\rm s}^{\rm ml}$ for various detection efficiencies ($\eta = 0.33,0.66,1.0$) and OPA gains ($g=0.75,1$).}
\end{figure}

In Fig.~\ref{fig-psig4} we show the photon-number statistics for a heralded field estimated to have $n_{\rm s}^{\rm ml}=5$ photons in it and compare it to a classical coherent state that has Poisson statistics with a mean number of photons equal to 5. From the figure it is clear that the photon-number distribution of the heralded state is much narrower than the distribution for the comparable classical coherent state. The width of the heralded field distribution is exactly the error in the maximum-likelihood estimate and is found using the mean squared error, i.e.~$(\Delta n_{\rm s}^{\rm ml})_{n_{\rm i}}^2 = \sum_{n_{\rm s}} (n_{\rm s}-n_{\rm s}^{\rm ml})^2 P_{\rm
sig}(n_{\rm s}|n_{\rm i})$.  Figure~\ref{fig-relerr} compares the ratio of the spreads of the heralded fields to their classical counterparts for different numbers of photons in the heralding field for various values of the single-photon detection efficiency $\eta$ and the OPA gain $g$. When this ratio is equal to unity, the heralded field has as much uncertainty in its number of photons as a Poissonian field. The smaller the ratio is, the closer the heralded field is to a Fock state.  The quality of the heralded state degrades as the loss increases and the gain of the OPA increases, as can be seen by noting that the ratio increases in these cases.

To gain further insight into the behavior of the heralded state,
we consider the mean number of photons in the heralded field. Although the mean number of photons cannot be an estimator of the number of photons in the heralded field, since it can take non-integer values, it is useful parameter that characterizes the field and can be analytically calculated. The mean number of photons in the signal field conditioned on $n_{\rm i}$ photons being detected in the idler field can be expressed in the following remarkably simple form (see Appendix)
\begin{equation}
    \label{eq-condmean}
    \langle n_{\rm s} \rangle_{n_{\rm i}} = \sum_{j=0}^{n_{\rm i}} a_j -1.
\end{equation}
Similarly, the conditional variance of the number of photons in the signal field can be expressed by
\begin{equation}
    \label{eq-condvar}
    (\Delta n_{\rm s})_{n_{\rm i}}^2  =  \sum_{j=0}^{n_{\rm i}} (a_j^2-a_j),
\end{equation}
where in both equations we have introduced the quantity $a_j = 
\left\{1-\tanh^2\!g\left[(1-\eta)+\eta(n-j)/M\right]\right\}^{-1}$ for notational convenience. An ideal system would produce Fock states with definite photon number so that $(\Delta n_{\rm s})_{n_{\rm i}}^2$ would vanish.  We note that $a_j$ cannot vanish since that requires $\eta$ to vanish, which is not a physically meaningful situation. Thus, $(\Delta n_{\rm s})_{n_{\rm i}}^2$ goes to zero only in the limit in which $a_j\rightarrow 1$. Also in this limit the conditional mean $\langle n_{\rm s}\rangle_{n_{\rm i}} \rightarrow n_{\rm i}$, which is consistent with the expected behavior of an ideal heralding system.

Two cases exist in which $a_j$ approaches unity.  The first case occurs in the low gain limit.  Irrespective of the system loss $\eta$, $a_j$ can be made arbitrarily close to one for small enough gain since $a_j < 1/(1-\tanh^2\!g) = 1+O(g^2)$. Intuitively, this limit is easily understood. When the gain is small enough, it is much more likely for the source to produce $n_{\rm s}$ photon pairs and the detector to count $n_{\rm i}$ of them than it is for the source to produce $n_{\rm s}+1$ pairs and the detector to still detect only $n_{\rm i}$. More quantitatively, there exists a gain such that the ratio $r = P_{\rm det}(n_{\rm i}|n_{\rm s}) P_{\rm OPA}(n_{\rm s})/(P_{\rm det}(n_{\rm i}|n_{\rm s}+1) P_{\rm OPA}(n_{\rm s}+1)) \gg 1$. This statement is easily verified by noting that $r \propto 1/\tanh^2\!g$ and can be made large for small enough $g$. The second case in which $a_j$ goes to one occurs when $M\gg n$ and $\eta = 1$.  In this limit the TMD behaves like an ideal photon number resolving detector and the system achieves perfect heralding.

Although these limiting cases are useful in providing insight into the behavior of the system, they are often not viable experimentally.  If the gain is too low, the rate at which the heralded states are produced becomes impractically small.  Furthermore in many cases $M$ cannot be too large, if for no other reason than that the loss incurred by adding splitters can become excessive. To treat the problem more generally, we calculate Mandel's $Q$ parameter defined by $Q = \left( (\Delta n_{\rm s})_{n_{\rm i}}^2-\langle n_{\rm s} \rangle_{n_{\rm i}}\right)/\langle n_{\rm s} \rangle_{n_{\rm i}}$~\cite{MandelWolfPoisson}. Mandel's $Q$ parameter is useful for characterizing the nature of the
photon-number distribution of a field.  The $Q$ parameter is bounded below by $-1$. In the limit in which $Q=-1$, the heralded field is in a Fock state. For $Q=0$, the field has Poissonian photon-number statistics; for positive and negative $Q$ the field has super- and sub-Poissonian statistics respectively. We use this parameter as a merit function describing the quality of the heralded states produced using the technique.
\begin{figure}
    \includegraphics[scale=0.65]{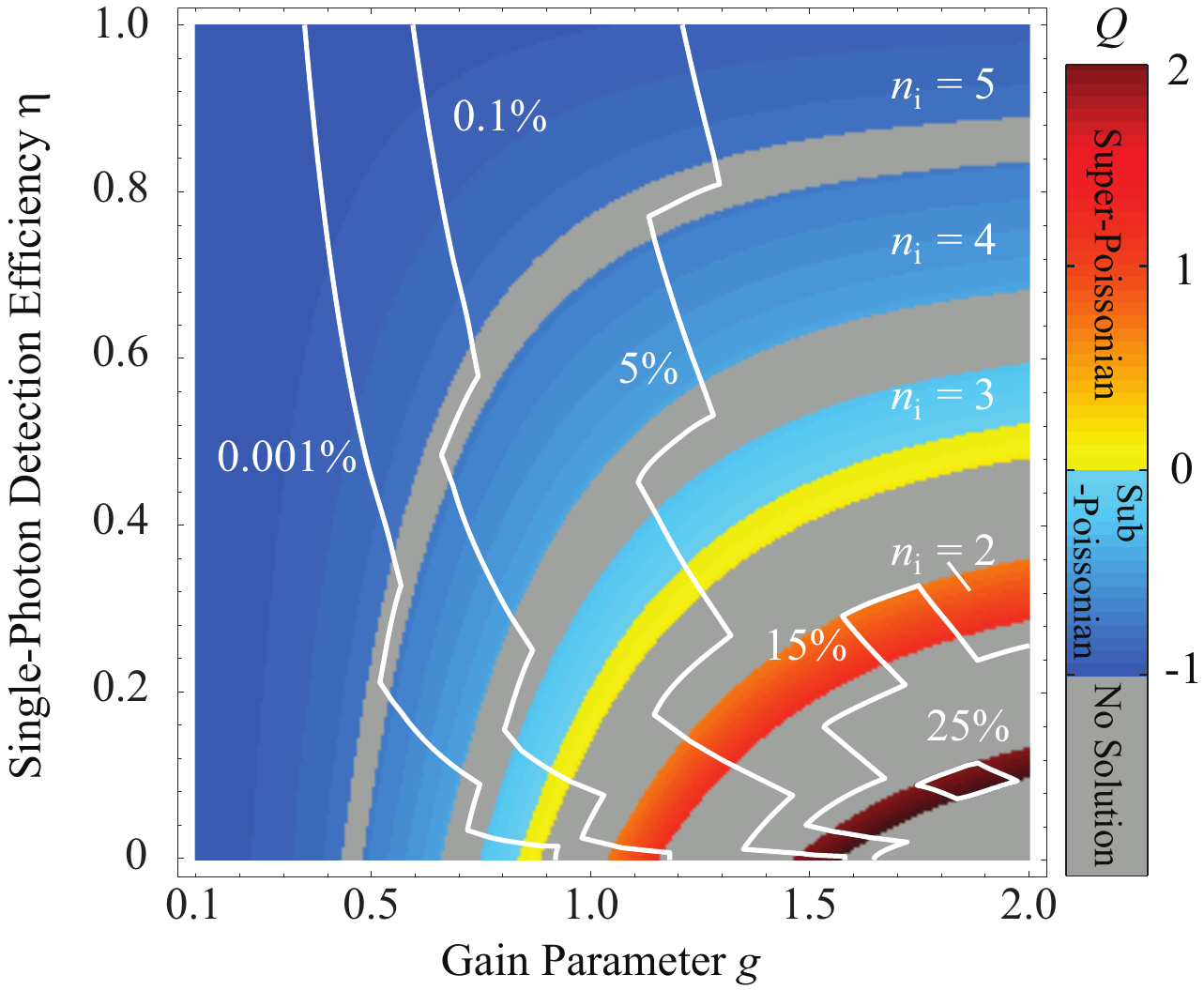}
    \caption{\label{fig-Q5} Mandel's $Q$ parameter versus the gain of the OPA and the single-photon detection efficiency of the time-multiplexed detector (TMD) for the case where $n_{\rm s}^{\rm ml} = 5$, i.e., the heralded field is a single-mode field estimated to contain 5 photons. The colored regions are labeled with the number of detected idler photons needed to herald the presence of 5 photons in the signal field.  The white contours are lines of constant $P(g, \eta; n_{\rm s}^{\rm ml})$, the probability of creating the heralded state.  The TMD is assumed to contain 5 beam splitters (M=32)}
\end{figure}

In practice, it is useful to know the TMD and OPA parameters necessary to produce a state that contains an estimated $n_{\rm s}^{\rm ml}$ photons and a certain value of the $Q$ parameter.  To this end, we calculate $Q$ at constant $n_{\rm s}^{\rm ml}$ versus the parameters $\eta$ and $g$.  Figure~\ref{fig-Q5} shows the result of the calculation for a heralding detector comprised of five beam splitters for the case when $n_{\rm s}^{\rm ml} = 5$. To calculate $Q$ for constant $n_{\rm s}^{\rm ml}$ it is necessary to invert Eq~(\ref{eq-Nml}) to find the appropriate value of $n_{\rm i}$ to be substituted into Eqs~(\ref{eq-condmean}) and (\ref{eq-condvar}). For certain values of the gain and detector efficiencies, Eq~(\ref{eq-Nml}) cannot be inverted, i.e.~there is no integer number of detected photons $n_i$ that correspond to a signal field in which $n_{\rm s}^{\rm ml} = 5$. In these instances, a signal field estimated to contain five photons is not possible to produce and is marked by the gray regions in the plot.  Also, the gray bands divide the space into regions corresponding to the number of detected idler photons that herald the presence of 5 photons in the signal field.  The figure confirms that the losses as well as the OPA gain should be kept as small as possible. In addition, the efficiency with which the heralded states are created is important.  The white contour lines overlaid on the plot show the probability with which the heralded state is successfully created, which is denoted by $P(g,\eta; n_{\rm_s}^{\rm ml})$. This probability is calculated from the sum $\sum_{k=0}^\infty P_{\rm det}(n_i|k)\,P_{\rm OPA}(k)$ where $n_i$ is obtained from the inversion of Eq.~(\ref{eq-Nml}) for different values of $g$ and $\eta$. The discrete jumps seen in the contour lines are due to the fact the $n_i$ can take only integer values. We see that it is possible to create nearly ideal Fock states even with significant detector loss; however, the efficiency of the creation of the heralded states is sacrificed.

Up to this point, we have considered only single-mode states produced by an OPA.  Although the single-mode limit can be reached, e.g.~by both spatially and frequency filtering~\cite{KobayashiPRA05}, the photons emitted by a free-space OPA are generally multimodal in nature and can be approximately described by a multimode thermal distribution~\cite{PaleariOE04}. The probability that the OPA will emit $k$ pairs of photons distributed among any $\mu$ modes is
given by
\begin{equation}
    \label{eq-pmu}
    P_{{\rm OPA}, \mu}(k) =
        \binom{k+\mu-1}{\mu-1}(1-\tanh^2\!g)^\mu\tanh^{2k}\!g,
\end{equation}
where $g$ is the parameter characterizing the gain per mode of the OPA.  We have assumed that each mode is equally likely to be occupied.  The multimode thermal distribution has a variance of $\langle k \rangle \left(1+\langle k \rangle/\mu\right)$. Since when $\langle k \rangle$ is held constant, the variance decreases with increasing number of modes, it is reasonable to expect that we may be able to create multimode fields containing near-definite numbers of photons under more relaxed conditions than by restricting ourselves to single-mode states.

\begin{figure}
    \includegraphics[scale=0.65]{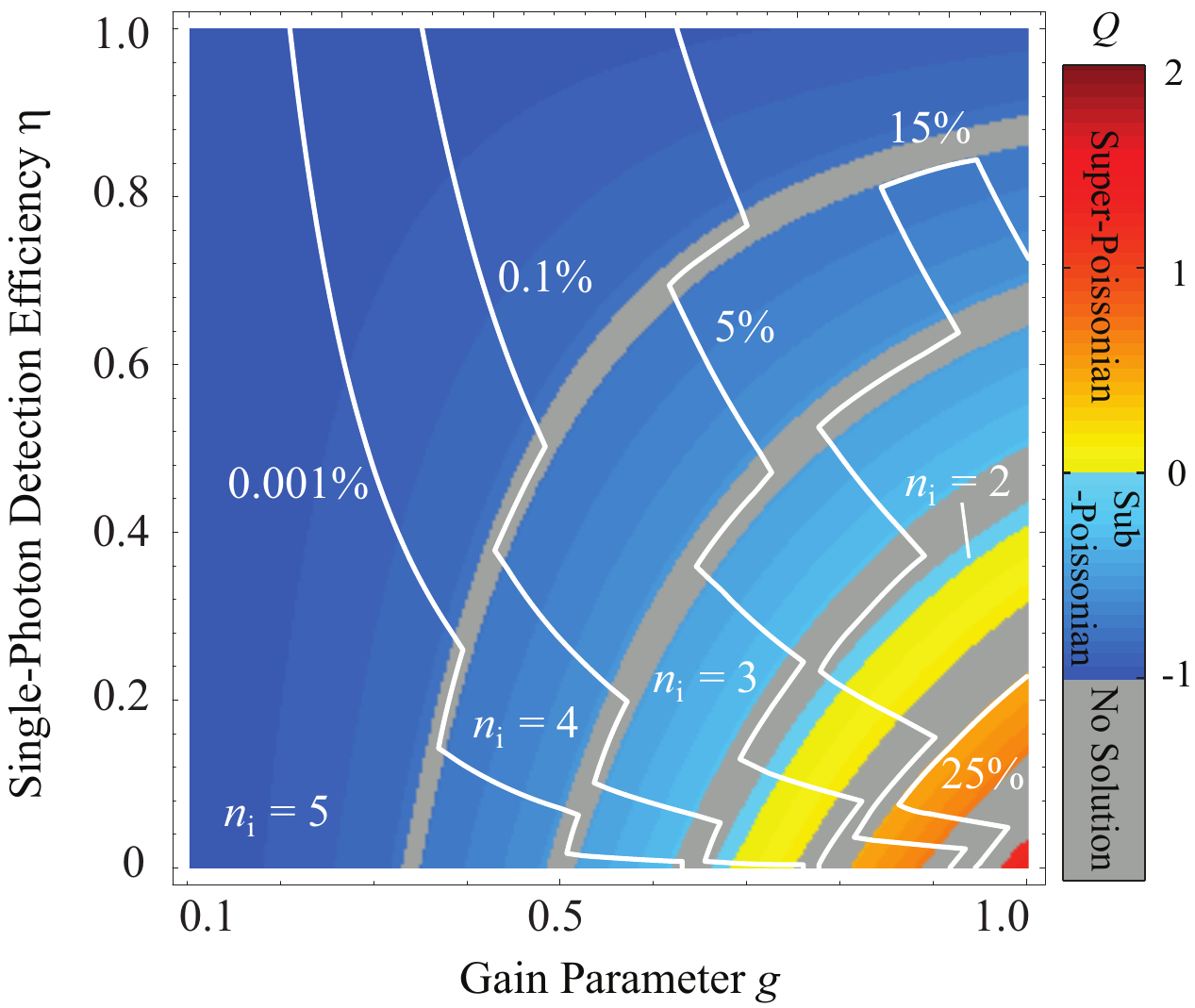}
    \caption{\label{fig-Q5a} Mandel's $Q$ parameter versus gain of the OPA and the single-photon detection efficiency of the TMD for the case where $n_{\rm s}^{\rm ml}=5$ and $\mu = 5$, i.e.~the signal field is estimated to contain 5 photons spread between 5 modes.  The colored regions are labeled with the number of detected idler photons needed to herald the presence of 5 photons in the signal field.  The white contours are lines of constant $P(g,\eta; n_{\rm s}^{\rm ml})$, the probability of creating the heralding state.}
\end{figure}

In Fig.~\ref{fig-Q5a} we plot the $Q$ parameter under the same conditions as in Fig.~\ref{fig-Q5} but now assuming that we are interested in the multimode output of the OPA (see Appendix for calculation details). We take $\mu =5$ as an example, which is typical value for a free space OPA~\cite{ParisPRL05}. When comparing Fig.~\ref{fig-Q5} and Fig.~\ref{fig-Q5a}, we see that the multimode case is more tolerant of loss in the heralding detector than the single mode case. For example, if we require a heralded state preparation rate of $5\%$, the heralded output of the OPA is sub-Poissonian so long as $\eta>0.11$, but in the single-mode case we have the more stringent requirement that $\eta>0.36$.

In conclusion, we have shown that it is possible to create multiphoton heralded states with highly non-classical photon-number distributions using a TMD as the heralding detector. We calculated Mandel's $Q$ parameter as a function of the detector loss and the OPA gain to characterize the quality of the heralded state.  Even in the presence of significant loss, we found that it is possible to produce states with highly sub-Poissonian photon-number distributions at reasonable efficiencies.  In addition, for situations where a single-mode output is not required, we showed that the full multimode output of an OPA can provide less uncertainty in the number of photons in the field than the single-mode output.  Since the TMDs involve only commercially available APDs and fibers, the technique we have presented here can be a particularly simple way to create states of light containing nearly definite numbers of photons.

\begin{acknowledgments}
The authors would like to thank Christine Silberhorn and Gerd
Leuchs for helpful discussions.  This work was supported by the US
Army Research Office through a MURI grant.  KWC thanks the support from the Croucher Foundation.
\end{acknowledgments}

\appendix*
\section{Appendix}
The derivation of the conditional mean and variance of the number of
photons in the signal field conditioned on $n_{\rm i}$ photons being
detected in the idler field is described in this appendix.

To calculate the conditional mean and variance, we need to
evaluate the sums defined by
\begin{equation}
    S_l = \sum_{n_{\rm s}=0}^\infty n_{\rm s}^{l} P_{\rm det}(n_{\rm i}|n_{\rm s}) P_{{\rm OPA},\mu}(n_{\rm s}), \quad
    l=0,1,2,
\end{equation}
where $P_{\rm det}(n_{\rm i}|n_{\rm s})$ and $P_{{\rm
OPA},\mu}(n_{\rm s})$ are the probability distributions given in
the text in Eq.~(\ref{eq-pngN}) and Eq.~(\ref{eq-pmu})
respectively.

The simplest case is that of $l=0$. Performing the summation over
$n_s$ by recognizing the negative binomial series, we find
\begin{equation}
    S_0 =
    A \sum_{j=0}^{n_i}\binom{n_{\rm i}}{j}\frac{(-1)^j}{(x+j)^\mu},
\end{equation}
where we have defined $A=\binom{M}{n_{\rm
i}}\left(M/(\eta\sinh^2\!g)\right)^\mu$ and $x=M/(\eta
\sinh^2\!g)+M-n_{\rm i}$ for simplicity. Next, using the identity
for Gauss hypergeometric series, $\sum_{j=0}^{n_{\rm i}}
\binom{n_{\rm i}}{j}(-1)^j/(x+j) = B(x,1+n_{\rm i})$, where
$B(a,b)\equiv \Gamma(a)\Gamma(b)/\Gamma(a+b)$ is the beta
function, we express $S_0$ as
\begin{equation}
    S_0 =
    A\frac{(-1)^{\mu-1}}{(\mu-1)!}\partial_x^{(\mu-1)}B(x,1+n_{\rm i}),
\end{equation}
where $\partial_x^{(\mu-1)}$ is shorthand for the $(\mu-1)$th-order
partial derivative with respect to $x$.

We follow an analogous procedure for the case when $l=1,2$, where
the primary difference is that the summations over $n_{\rm s}$ are
now derivatives of the negative binomial series.  The resulting
sum $S_1$ is given by
\begin{equation}
S_1 =
A\frac{\mu(-1)^\mu}{\mu!}\partial_x^{(\mu)}\left[\left(x+b\right)B(x,1+n_{\rm
i})\right],
\end{equation}
with  $b=M(1-\eta)/\eta+n_{\rm i}$.  The sum $S_2$ is similarly
given by
\begin{equation}
    \begin{split}
        S_2 = A\frac{\mu^2(-1)^{\mu+1}}{(\mu+1)!}\partial_x^{(\mu+1)}\left[\left(x+b\right)^2 B(x,1+n_{\rm i})\right. \\
         \left.+ \frac{(x+b)c}{\mu}B(x,1+n_{\rm i})\right],
    \end{split}
\end{equation}
where $c=M/(\eta \tanh^2\!g)$ is defined for simplicity.

The conditional mean and variance can be expressed in terms of
these sums as $\langle n_{\rm s} \rangle_{n_{\rm i}} = S_1/S_0$
and $(\Delta n_{\rm s})_{n_{\rm i}}^2 = S_2/S_0-(S_1/S_0)^2$.
Substituting into these equations and simplifying, we find the
general form for the conditional mean and variance to be
\begin{eqnarray}
    \langle n_{\rm s} \rangle_{n_{\rm i}} &=& -\mu-c\, g_\mu(x)\\
    (\Delta n_{\rm s})_{n_{\rm i}}^2 & =
    &\left(c+c^2 \partial_x\right) g_\mu(x),
\end{eqnarray}\\
where $g_{\mu}(x) =
\partial_x^{(\mu)}B(x,1+n_{\rm i})/\partial_x^{(\mu-1)}B(x,1+n_{\rm i})$.  By setting $\mu=1$ it is not difficult to recover the single-mode
results of Eq.~(\ref{eq-condmean}) and Eq.~(\ref{eq-condvar})
shown in the text.  Furthermore, the single-mode and multimode
expressions for $Q$ can be easily found from these results.

\end{document}